\input{aipcheck}

\documentclass[
    ,final            
  ]
  {aipproc}

\layoutstyle{8x11single}

\begin{document}

\title{Charge Symmetry Breaking and Parity Violating Electron Scattering}

                

\author{Gerald A. Miller}{
  address={Physics Department, University of Washington, Seattle WA 98195-1560}}

\begin{abstract}
 I review the effects of charge symmetry breaking CSB  on electromagnetic form factors
 and how that influences extraction of information regarding nucleon strangeness content and
 the weak mixing angle. It seems that  CSB effects are  very modest and should not impact 
 the analysis of experiments.  \end{abstract}

\maketitle


\section{Introduction}

Charge symmetry CS  is invariance under an isospin  rotation of $\pi$ about the $y$-axis in isospin space. Thus a $u$ 
  quark is rotated into a $d$ quark. CS is broken slightly by the light-quark mass difference and by electromagnetic effects.
  Isospin invariance, or $[H,T_i]=0$ is invariance under all rotations in isospin space. This invariance  is also called charge independence, CI. Charge symmetry does not imply isospin invariance. Diverse aspects
  of charge symmetry and its breaking have been reviewed~\cite{Henley:1979ig}-
  \cite{Miller:2006tv}.

  For example, the mass difference between charged and neutral pions breaks isospin invariance but not charge symmetry. Another example is the difference between the $np$ and $nn$ scattering
  lengths. This difference does not involve isospin mixing between $ T=0 $ and $T=1$ states.
  
  In general the size of CSB effects  is much smaller than the breaking of isospin invariance, CIB. The 
  scale of CSB is typified by the ratio of the neutron-proton mass difference to the proton mass which is about one part in 1000. This is much smaller than the pion mass difference effect which is one part in 27. The CIB of nucleon-nucleon scattering lengths was discovered well before 1965 but the measurement of their  CSB of had to wait until about  1979.
  Thus the expectation is that CSB is a small effect, uncovered only with special effort. The small relative  size of CSB effects compared with those of CIB is a consistent with the power counting of
  of chiral perturbation theory~\cite{vanKolck:1996rm}.

\section{Parity Violating Electron Scattering, Strangeness Electromagnetic Nucleon Form Factors and Charge Symmetry Breaking}
 Understanding parity-violating electron scattering requires knowledge of weak neutral form factors. These are sensitive to nucleon strangeness content and
 the value of the weak mixing angle. The latest review is that of Armstrong \& McKeown~\cite{Armstrong:2012bi} who conclude that a  convincing signal for nucleon strangeness has not been seen.

 
 The nucleon Dirac $F_1$ and Pauli $F_2$ electromagnetic form factors can be composed according to their quark content:
 \begin{eqnarray}
  F_{1,2}^\gamma = \frac{2}{3} F_{1,2}^u - \frac{1}{3} F_1^d - \frac{1}{3} F_{1,2}^s.\;
\end{eqnarray}
If one considers the elastic interaction of the $Z$-boson with the proton 
and  uses
charge  symmetry  ($u$ in proton = $d$ in neutron, $d$ in proton = $u$ in neutron)  to relate $F_{1,2}^Z$ to measured form factors one finds
\begin{equation}
 G_{E,M}^{Z,p} = (1 - 4 \sin^2 \theta_W) G_{E,M}^{\gamma,p}
		- G_{E,M}^{\gamma,n} - G_{E,M}^{s}, \label{eq:EMZ}
\end{equation}
if the result is expressed in terms of Sachs form factors.

\begin{figure}
\includegraphics[height=.2\textheight]{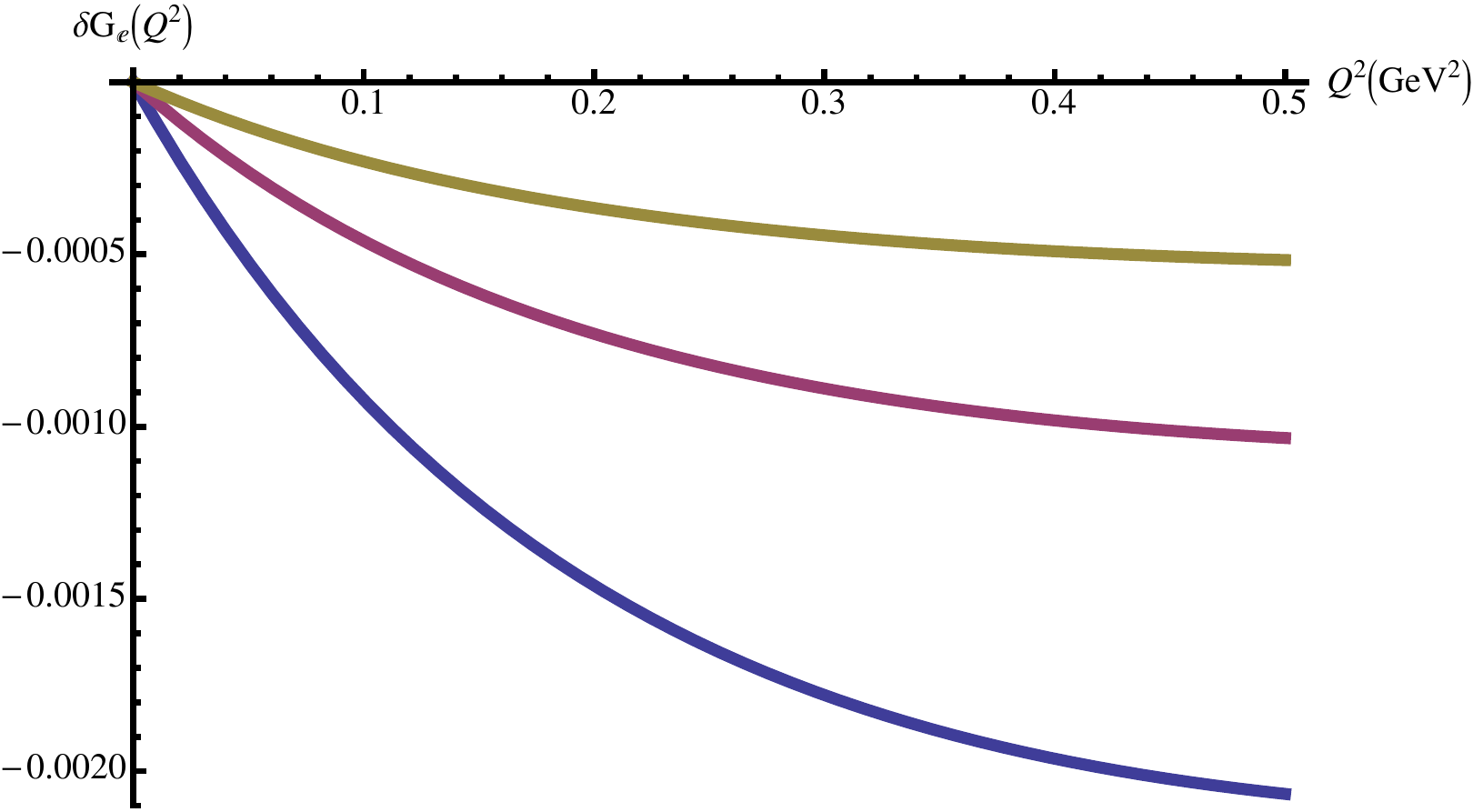}
\includegraphics[height=.2\textheight]{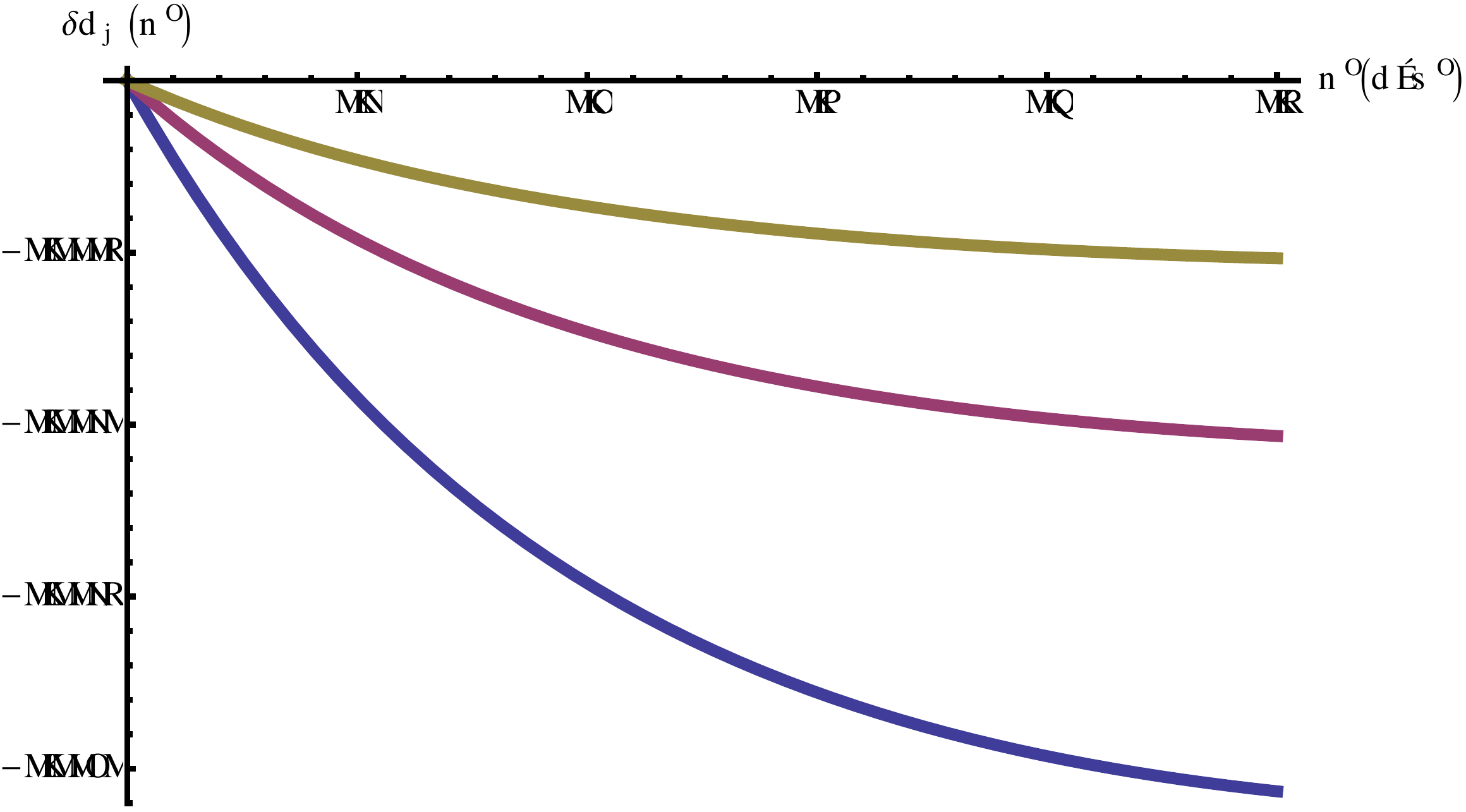}
  \caption{(Color Online)  The change in electric $\delta G_E$ and magnetic $\delta G_M$ form factors due to the effects of CSB~\cite{Miller:1997ya} . The three non-relativistic models are defined by the 
fraction (0.8 (blue), 0.67 (red) , 0.33( green)) that one gluon exchange contributes to 
of $\Delta N$ Splitting. The computed value of $M_n-M_p$ is the same in each model.}
\end{figure}

Including the effect of CSB changes the above equations. The matrix element of the $Z$-boson nucleon interaction can be written
\begin{equation}
 Z_\mu =\bigl\langle p(\vec p\,')\,\bigl|\,(1 - \frac{8}{3} \sin^2 \theta_W) j^u_\mu  + (-1+ \frac{4}{3} \sin^2 \theta_W) (j_\mu^d +j_\mu^s)
  \,\bigr|\,p(\vec p)\bigr\rangle
\end{equation}
We need to relate $\bigl\langle p(\vec p\,')\,\bigl| j^{u,d}_\mu \,\bigr|\,p(\vec p)\bigr\rangle $ to measured form factors. To do this we decompose the physical  nucleon wave functions into pieces that respect CS and those that do not. Thus we write
\newcommand{\bea}{\begin{eqnarray}}
\newcommand{\eea}{\end{eqnarray}}
\bea
\bigr|\,p(\vec p)\bigr\rangle=\bigr|\,p_0(\vec p)\bigr\rangle+\bigr|\,\Delta p(\vec p)\bigr\rangle,\,\,
\bigr|\,n(\vec p)\bigr\rangle=\bigr|\,n_0(\vec p)\bigr\rangle+\bigr|\,\Delta n(\vec p)\bigr\rangle
,\eea
where $\bigl|\Delta p(\vec p)\,\bigr\rangle,\,\bigl|\Delta n(\vec p)\,\bigr\rangle,$ is caused by the CSB part of the Hamiltonian: $\Delta H\equiv [H,P_{\rm cs}]$. Then
\begin{equation}
 G_{E,M}^{Z,p} = (1 - 4 \sin^2 \theta_W) G_{E,M}^{\gamma,p}
		- G_{E,M}^{\gamma,n} - G_{E,M}^{s} +\bigl\langle p_0(\vec p\,')\bigl|\,{2\over3}j^d_\mu-{1\over3}j^u_\mu\bigr|\Delta p(\vec p)\bigr\rangle_{E,M}
+\bigl\langle \Delta p(\vec p\,')\bigl|\,{2\over3}j^d_\mu-{1\over3}j^u_\mu \bigr|p_0(\vec p)\bigr\rangle_{E,M}\label{csb}
\end{equation}
where the subscript $E,M$ refers to taking the matrix element that leads to the electric or magnetic form factor. Our interest here is in   the size and uncertainty in our knowledge of the  last two terms on the  right-hand side of Eq.~(\ref{csb}).
 
Experimentalists currently believe that  the uncertainty attached to charge symmetry is now limiting the ability to push further on the strange form factors. We want to learn if uncertainty in CS and its breaking really limits the ability to push further. 
\section{My 1998 paper}

I studied the CSB of electromagnetic form factors in~\cite{Miller:1997ya}.
 Using a set of $SU(6)$ non-relativistic quark models, the effects of the charge symmetry breaking Hamiltonian were considered for experimentally relevant values of the momentum transfer and found to be less than about 1\%. The percentage is relative to the values of $G_{E,M}$.  The size of current experimental error bars, typified by a magnitude of about 0.01  is more relevant.

CSB induces  changes of proton electromagnetic form factors $\delta G_{E,M}$. The use of perturbation theory gives 
 \bea \delta G_E(Q^2=\vec{q}\,^2)
 =\bigl\langle p_0\bigl|\,\sum_{i=1}^3 \left({1\over3}+\tau_3(i)\right)
 e^{i\vec{q}\cdot\vec{r}}{\Lambda\over M_p-H_0}2\Delta H\,\bigr|p_0
 \rangle,\\
 \delta G_M(Q^2=\vec{q}\,^2)
 =\bigl\langle p_0\bigl|\,\sum_{i=1}^3  
  {{\sigma}_3(i)\over 2m_q}
 e^{i\vec{q}\cdot\vec{r}}{\Lambda\over M_p-H_0}2\Delta H\,\bigr|p_0\rangle.\eea
 to first-order in CSB effects.
Our $\Delta H$  was derived by using the light quark mass difference  $m_d-m_u$ in the kinetic energy and one gluon exchange operators, and the effects of   one photon exchange.
The operator 
$\Lambda$ projects out of ground state. Thus  if $\vec q$=0, $\delta G_{E,M}(0)=0$
because of orthogonality and because  the operator $\Delta H$ does not excite the $\Delta$.
Therefore, in this approach, the effects of  CSB must be ignorable   if $Q^2$ takes on    small  values. This is shown in  Fig.~1. 

\section{Paper of Kubis \& Lewis~\cite{Kubis:2006cy}
~-What I omitted  }

Kubis \& Lewis KL found terms that I missed in the pion cloud of the nucleon~Fig.~2.  The CSB effect occurs in the mass difference of intermediate nucleon states. This small  mass difference does not lead to small csb effects
because  of a logarithmic divergence in the loop diagram that arises from the use of heavy baryon effective field theory. KL  remove the divergence using a resonance saturation procedure in which the upper limit of the logarithm is evaluated at the rho meson mass and  the effects of $\rho^0-\omega$ mixing, Fig. 3, are included. The result of their calculation is that $\delta G_M$ can be characterized as $\sim 0.02\pm 0.02$ for values of $Q^2$ between 0 and 0.3 GeV$^2$~\cite{Kubis:2006cy}. Such a large range indeed hinders the extraction of strangeness content or the value of the weak mixing angle. 
\begin{figure}
\includegraphics[height=.14\textheight]{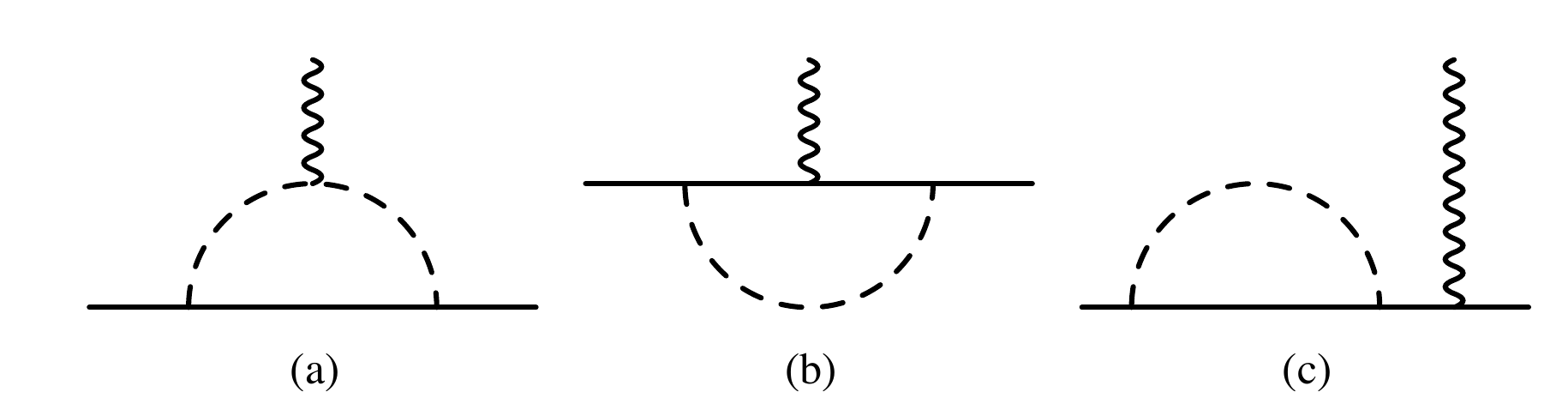}
  \caption{ Pion loop graphs of Kubis \& Lewis. The CSB effect occurs in the mass difference of intermediate nucleon states.}
\end{figure}
\begin{figure}
\includegraphics[height=.14\textheight]{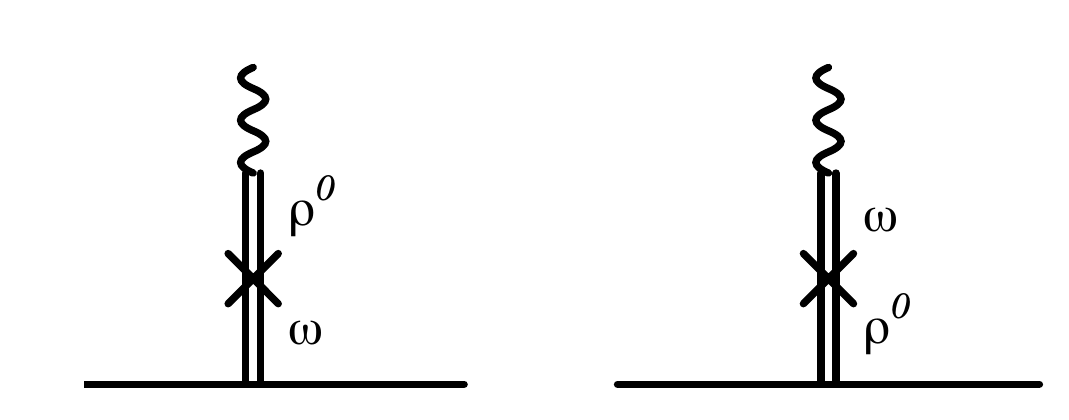}
  \caption{ Resonance saturation  graphs of Kubis \& Lewis. The CSB effect occurs in the conversions $\rho\leftrightarrow\omega$.}
\end{figure}

In the KL calculation the large uncertainty is driven by uncertainty in the strong coupling constants used. 
KL take strong coupling constants from dispersion analyses of electromagnetic form factors. A large spread in the $\omega$-nucleon  tensor coupling is obtained, and the
vector  coupling constant 
is much larger  $g_\omega^2/4\pi\sim150$ than typically used in nucleon-nucleon  scattering $g_\omega^2/4\pi\sim10-20.$ Is there a
scientific way to choose between sets of strong coupling constants extracted from different processes?

 \section{$\rho^0-\omega$ mixing in NN scattering}

The nucleon-nucleon potential obtains a CSB contribution when in its flight from one nucleon to another a $\rho^0$ meson converts to an $\omega$ 
meson~\cite{Henley:1979ig}-
\cite{Miller:2006tv}.  This effect gives rise to a medium ranged potential which  is of the correct sign and magnitude to explain the difference between
the $np$ and $nn\;^1S_0$ scattering lengths. When such a potential is used in nuclear physics it can account for the Nolen-Schiffer 
anomaly~\cite{Coon:1987kt,Blunden:1987wr}. This effect also yields a CSB Class IV  (spin-orbit) force that was observed in pioneering experiments at TRIUMF and IUCF that compared the analyzing powers of neutrons and protons. See the review~\cite{Miller:1994zh}.

The use of the KL parameters implies a huge range in the values of the CSB potential in the $1S_0$ state. The use of such a range leads to CSB effects in severe disagreement with bound-state and scattering data.

\section{Another way to deal with short distance uncertainty-relativistic chiral perturbation theory}
One does not have to use the heavy baryon expansion. One may make relativistic calculations~\cite{Fuchs:2003qc} to implement chiral perturbation theory.
 In that case the CSB  contributions of the graphs are finite.
One can go further by 
taking the  intermediate nucleon to be on mass-shell, and then using  known $\pi-N$  form factors~\cite{Alberg:2012wr}.  
Preliminary results of such calculations show that the contributions to $\delta G_M$ are about 0.01 (with a much smaller uncertainty) at small values of $Q^2$.
As expected, the effect is of order $(M_n-M_p)/M_p$  times the contribution of the pion loop graph to the proton anomalous magnetic moment.

\section{Summary}
I used 
quark {\it model} KL used   chiral perturbation {\it theory}, with the presumption that theory is better than a model.
However,  if an unconstrained counter term is needed to evaluate the theory, the theory becomes a  model.  More constraints must be implemented
to improve either.

I found small $<0.002$ CSB effects on electromagnetic form factors in 1998
Kubis \& Lewis found a larger uncertainty of CSB about $\sim 0.02\pm 0.02$ for $\delta G_M$.
However,  CSB in NN scattering constrains the strong coupling constants  needed for KL to evaluate the effects of resonance saturation and significantly reduce the
CSB effects in the proton electromagnetic form factors. The 
actual size of CSB effect is probably pretty small, of the order of 0.1\% of the anomalous magnetic moment. I hope to present more precise numbers soon.
\begin{theacknowledgments}
   This   work has been partially supported by 
U.S. D. O. E.  Grant No. DE-FG02-97ER-41014
\end{theacknowledgments}

\bibliographystyle{aipproc}

\end{document}